\documentclass[twoside,twocolumn,a4paper]{IEEEtran}
\usepackage{url}
\usepackage{mathtools}
\usepackage{amsfonts}
\usepackage{setspace}
\usepackage{mathptmx}       
\usepackage{helvet}         
\usepackage{courier}       
\usepackage{type1cm}       
\usepackage{makeidx}
\usepackage[bottom]{footmisc}
\usepackage{multicol}
\usepackage{amsmath}

\begin{document}

\title{Comment on `On the Quantum Theory of Molecules' [J. Chem.Phys. {\bf 137}, 22A544 (2012)]}
\author {Brian T. Sutcliffe\thanks{B.T. Sutcliffe, Service de Chimie quantique et Photophysique,  Universit\'{e} Libre de Bruxelles, B-1050 Bruxelles, Belgium. {email: bsutclif@ulb.ac.be}},~~ R. Guy Woolley\thanks{R.G.Woolley, School of Science and Technology, Nottingham Trent University, Nottingham NG11 8NS, U.K.} 
\thanks{Manuscript accepted by \textit{Journal of Chemical Physics}, December 2013.}}
\maketitle

\begin{abstract}
In our previous paper [J. Chem.Phys. {\bf 137}, 22A544 (2012)] we argued that the Born-Oppenheimer approximation could not be based on an exact transformation of the molecular Schr\"{o}dinger equation. In this Comment we suggest that the fundamental reason for the approximate nature of the Born-Oppenheimer model is the lack of a complete set of functions for the electronic space, and the need to describe the continuous spectrum using spectral projection.
\end{abstract}

After removal of the centre-of-mass variables, the internal molecular Hamiltonian may be written as\cite{SW:12, SW:13}
\begin{equation}
\mathsf{H}'~=~\mathsf{T}_{Nu}~+~\mathsf{V}
\label{Moldec}
\end{equation}
where $\mathsf{T}_{Nu}$ is the nuclear kinetic energy. In the Born-Huang method $\mathsf{V}$ is identified with the electronic Hamiltonian for stationary nuclei (the `clamped-nuclei' Hamiltonian). Let $X$ stand for the nuclear positions, and $x$ for the electronic positions. Then for fixed $X_f$, Born-Huang write\cite{MB:51,BH:54}
\begin{equation}
\mathsf{V}~=~\mathsf{H}_o(X_f)
\label{cnHam}
\end{equation}
where $\mathsf{H}_o(X_f)$ is an operator on the electronic Hilbert space, $L^{2}_{x}$, with `eigenfunctions' \{$|\varphi_j(x,X)\rangle$\}, 
(where $j$ stands for both discrete and continuous labels), supposed `complete' in $L^{2}_{x}$ so that there is a resolution of the identity, at fixed $X_f$,
\begin{equation}
\mathsf{I}(X_f)_x~=~\sum_{j}|\varphi_j(x,X_f)\rangle \langle \varphi_{j}(x,X_f)|.
\label{idens}
\end{equation}

The Born-Huang method has three steps: firstly use $\mathsf{I}(X_f)_x$ to write an expansion of a molecular wavefunction, $|\Psi_{i}\rangle$,
\begin{eqnarray}
|\Psi_{i}\rangle~&=&~\sum_{j}|\varphi_j(x,X_f)\rangle \langle \varphi_{j}(x,X_f)|\Psi_{i}\rangle_x\\~&\equiv&~\sum_{j}a_{ij}(X_f)|\varphi_{j}(x,X_f)\rangle.
\label{expans}
\end{eqnarray}

Then write $\mathsf{H}'$ as in equation (\ref{Moldec}) with $\mathsf{V}$ identified by (\ref{cnHam}) and apply it to the expansion of molecular states,
\begin{equation}
0~=~(\mathsf{H}'-E_i)\sum_{j}|\varphi_{j}(x,X_f)\rangle~a_{ij}(X_f).
\label{expansBH}
\end{equation}
Finally in the third step form the scalar product in $\mathsf{L}^{2}_{x}$ of this expression with a `basis' vector, $|\varphi_{k}(x,X_f)\rangle$
\begin{equation}
\langle \varphi_{k}(x,X_f)(\mathsf{H}'-E_i)|\sum_{j}|\varphi_{j}(x,X_f)\rangle_x~a_{ij}(X_f)=0.
\label{aeqns}
\end{equation}
This leads in the well-known way to a system of coupled differential equations for the coefficients \{$a_{ij}(X)$ which are the unknowns - the nuclear wavefunctions. In the Chemistry and Physics literature this is commonly presented as an exact transformation of the original Schr\"{o}dinger equation \cite{AMG:12,LSC:13}.

In our earlier paper \cite{SW:12} we argued that for the decomposition (\ref{Moldec}) to be valid, the operator $\mathsf{V}$, which we denoted as $\mathsf{H}^{\mbox{elec}}$, must be written as a \textit{direct integral} over clamped-nuclei electronic Hamiltonians, the integral being taken over all nuclear positions; we further suggested that after this correction an expansion analogous to (\ref{expans}) would be problematic. The operator $\mathsf{H}^{\mbox{elec}}$ has purely continuous spectrum extending from some minimum value to $+\infty$ on the real-axis; it has no true eigenvalues, and no normalizable eigenvectors in the Hilbert space $L^2(x,X)$. We emphasize that this description is entirely in agreement with the mathematical physics literature but not with the conventional Born-Huang discussion summarized above since the distinction between $\mathsf{H}_o(X_f)$ and its direct integral $\mathsf{H}^{\mbox{elec}}$ is obvious and fundamental.

The recent work of Jecko \cite{TJ:13} gives a critical mathematical summary of the Born-Oppenheimer approximation including the expansion approach described by (\ref{idens}) - (\ref{aeqns}), and emphasizes the fundamental role of spectral projection for operators with continuous spectra. Thus it is very much to the point that there is no `complete set of eigenfunctions' in Hilbert space for the clamped-nuclei Hamiltonian $\mathsf{H}_o(X_f)$ required for the expansion (\ref{expans}) (and likewise for $\mathsf{H}^{\mbox{elec}}$). This is crucial since the Born-Huang method effectively writes the clamped nuclei Hamiltonian in a spectral form using the `eigenfunctions' to provide a resolution of the identity
\begin{equation}
\mathsf{H}_o(x,X_f)~=~\sum_{j}E_{j}(X_f)|\varphi_{j}(x,X_f)\rangle \langle \varphi_{j}(x,X_f)|
\end{equation}
Such a representation of an operator is valid for self-adjoint operators on a finite dimensional Hilbert space, or for operators with \textit{purely discrete spectrum} on an infinite dimensional Hilbert space; thus it would be valid for a system of light and heavy coupled oscillators which is often used as a model for testing the Born-Oppenheimer approximation. But it is not generally valid for unbounded operators with continuous spectra\cite{JMJ:72, TK:80}, as is the case here.

One might hope that $\mathsf{V}$ (either choice) has `generalized eigenfunctions' satisfying
\begin{equation}
\mathsf{V}|\varphi_{E,n}\rangle~=~E|\varphi_{E,n} \rangle
\end{equation}
where
\begin{eqnarray}
\langle \varphi_{E'',n} | \mathsf{V}|\varphi_{E',m}\rangle~&=&~E'~\delta(E''~-~E')~\delta_{nm}\nonumber\\
\langle \varphi_{E'',n}|\varphi_{E',m}\rangle~&=&~\delta(E''~-~E')~\delta_{nm}
\label{4}
\end{eqnarray} 
However nothing is known about the properties of such `generalized eigenfunctions' for Hamiltonians with Coulombic interactions other than for the special case of the 2-body system (H atom); even their very \textit{existence} does not seem to be demonstrated for $\mathsf{V}$ derived from the general $N$-particle Coulomb Hamiltonian\cite{TJ:13}. Thus there is no way to check an expansion of the molecular wavefunction based on a set of such `generalized eigenfunctions' (for either choice of $\mathsf{V}$), and such an approach cannot  be described rationally as `exact'.

To deal correctly with the continuous spectrum we need the following abstract result; to functions of a self-adjoint operator $\mathsf{L}$ one can associate a projection operator valued function (measure) $\mathsf{P}_{\lambda}$ through the spectral relation \cite{JMJ:72,TK:80,RS1:75}
 \begin{equation}
   \mathsf{f(L)} = \int\limits_{-\infty}^{+\infty} f(\lambda)~ d\mathsf{P}_{\lambda}.
 \end{equation} 
where the spectral projector  $\mathsf{P}_{\lambda}$ is given formally by the Stone formula, which relates functions of a self-adjoint operator to the discontinuity of its resolvent across the real axis \cite{RS1:75}. The spectral theorem guarantees that the spectral family $\mathsf{P}_{\lambda}$ is orthogonal, `diagonalizes' the operator $\mathsf{L}$ and provides a resolution of the identity. Spectral projection replaces the notion of `complete set of states' which is valid in general only for operators with purely discrete spectra; there is no general practical method for obtaining a family of spectral projectors and we do not propose this as an alternative route for molecular theory. These formal results do however justify our earlier claim that it is not possible to reduce the molecular Schr\"{o}dinger equation to a system of coupled differential equations of classical type for nuclear motion on PES, as suggested by Born\cite{MB:51}, without a further approximation of an essentially empirical character\cite{SW:12}. 

The crucial point emphasized by Jecko\cite{TJ:13} is that if one gives up the goal of an exact computation of the molecular wavefunction $|\Psi\rangle$ one still has the opportunity to use ideas about PES to seek a \textit{good approximation} to it. The essential step is to fix the total energy of the molecular system at the outset and then construct an effective Hamiltonian depending on this energy working with the purely discrete part of the spectrum. The accuracy of the Born-Oppenheimer approximation is then defined by the quality of the replacement of the true Hamiltonian by the effective one (the `adiabatic Hamiltonian') and is measured in terms of the usual small parameter related to the electron/nucleon mass ratio. The procedure turns out to be universal in the sense that it applies to bound states, resonances, non-resonant scattering and time evolution, and it avoids the construction of the family of spectral projectors required for an exact description of the continuous spectrum. Mathematically it is an example of \textit{semiclassical analysis}\cite{TJ:13}. 

The clamped-nuclei Hamiltonian and the associated notion of PES are
crucial to the method; on the other hand PES make no appearance in the
formally exact description based on spectral projection, so it is hard
to claim any fundamental role for Potential Energy Surfaces in the
quantum mechanics of molecules. This is not of course, to ignore the
important role of PES in approximate calculations and their role in
interpreting experimental results. It is simply to place PES properly
in relation to quantum theory.

We are aware that the above considerations have a much wider range of applicability in Chemistry and Physics than simply the Born-Oppenheimer approximation, for example the conventional perturbation theoretic accounts of the electrical, magnetic and optical properties of substances commonly involve sums over `complete sets of molecular states'. It seems likely that the strategies to avoid spectral projection (similar to that described by Jecko\cite{TJ:13}) to be found in the book by Kato\cite{TK:80} are relevant here.

\section*{Acknowledgement}
It is a pleasure to acknowledge stimulating discussions with Dr. T. Jecko.

\end{document}